\documentstyle[11pt,aaspp4]{article}

\def\simg{\mathrel{\hbox{\rlap{\lower.55ex \hbox {$\sim$}}
                   \kern-.3em \raise.4ex \hbox{$>$}}}}
\def\siml{\mathrel{\hbox{\rlap{\lower.55ex \hbox {$\sim$}}
                   \kern-.3em \raise.4ex \hbox{$<$}}}}
\def\Mesz{M\'esz\'aros~}
\def\refe{\reference}
\def\Pacz{Paczy\'nski~}
\def\beq{\begin{equation}}
\def\enq{\end{equation}}
\def\bea{\begin{eqnarray}}
\def\ena{\end{eqnarray}}

\def\bec{\begin{center}}
\def\enc{\end{center}}
\def\etal{{\it et al.}}
\def\ergs{\hbox{erg s$^{-1}$}}
\def\cmcui{{\rm cm}^{-3}}
\def\r13{r_{13}}
\def\T8{T_8}
\def\L47{L_{47}}
\def\Lm{L_{m}}
\def\E53{E_{53}}

\def\msun{M_\odot}

\begin{document}

{\it \hfill ApJL, in press}
%{\it \hfill sub. 7/30/00, acc. 10/13/00}\\

\title{ Fe K-$\alpha$ Emission from a Decaying Magnetar Model \\
  of Gamma-Ray Bursts}

\author{M.J. Rees $^1$ \& P. \Mesz$^{1,2,3}$}
\smallskip\noindent
$^1${Institute of Astronomy, University of Cambridge, Madingley Road, Cambridge
CB3 0HA, U.K.}\\
\smallskip\noindent
$^2${Dpt. of Astronomy \& Astrophysics, Pennsylvania State University,
University Park, PA 16803}\\
$^3${Institute for Theoretical Physics, University of California, 
Santa Barbara, CA 93106-4030}

%\bec Draft :~~{\today} \enc

\begin{abstract}
The recent report of X-ray Fe features in the afterglow of the gamma-ray burst
GRB 991216 may provide important clues for identifying the nature of its
progenitor and constraining the burst mechanism. We argue that the strong
line emission can be attributed to the interaction of a continuing (but
decaying) post-burst relativistic outflow from the central engine
with the progenitor stellar envelope at distances less than a light-hour.
Only a small  mass of Fe is then required, which could have been readily
produced by the star itself.
\end{abstract}


\section{Introduction}

Recently, Fe K-$\alpha$ and K-edge X-ray features have been reported with high
significance in the afterglow of the bright burst GRB 991216 and GRB 000214
after about 1 day (Piro, \etal, 2000, Antonelli et al, 2000).  
A straightforward interpretation of this observation would imply a mass 
$\simg 10^{-2}-1 \msun$ of Fe at a distance of about one light-day, 
possibly due to a remnant of an explosive event or supernova which
occurred days or weeks prior to the gamma-ray burst itself.

Here, we suggest an alternative and perhaps less restrictive scenario:  an
extended, possibly magnetically dominated wind from a GRB impacting the
expanding envelope of a massive progenitor star. This could be due either to a
spinning-down millisecond super-pulsar or to a highly-magnetised  torus around a
black hole (e.g. Wheeler, \etal 2000), which could produce a luminosity that was
still, one day after the original explosion,  as high as $L\sim 10^{47}\ergs$.
This luminosity may not dominate the continuum afterglow; but we argue that  it
could be efficiently reprocessed, by a modest amount $M_{Fe} \sim 10^{-8}\msun$
of material at  distances $\siml 10^{13}$ cm, into an Fe line luminosity
comparable to the observed value, together with a contribution to the X-ray
continuum.  Under this interpretation, the dominant continuum flux in the
afterglow, even in the X-ray band, is still attributable to a standard
decelerating blast wave.

\section{Energy Input and Fe Line Strength}

The typical GRB model assumes that the energy input episode is brief, typically
$t_{b} \siml 1-10^2$ s,  its energy and mass outflow being either a delta or
a top-hat function. However, peculiarities in the early stages of some
afterglows, e.g GRB 970508, have served as motivation for considering a more
extended input period in which the energy or mass outflow rate may vary in
time and the late energy input could exceed the prompt contribution (e.g. Rees
\& \Mesz 1998, Sari \& \Mesz 2000).  Here we consider a related model, in
which the power output continues at a diminishing rate, for a longer time of
hours to days. The prolonged  activity could arise if orbiting debris around a
newly-formed black hole takes a long time to be completely swallowed, or if the
central object becomes a fast-spinning neutron star, rather than a black hole.
Collapsar or hypernova models (e.g. MacFadyen \& Woosley, 1999; \Pacz, 1998;
Woosley 1993) or magnetar-like GRB models (Usov, 1994, Thompson, 1994, Wheeler
\etal~ 2000) provide a natural scenario for a sudden burst followed by a more
slowly decaying energy input.

The power output would be primarily in a magnetically-driven relativistic wind
(which, even during the later phases discussed here, would be hugely 
super-Eddington). The generic magnetized outflow from a spinning compact 
object is
\beq
\Lm \sim  2\pi r_o^2 c (B^2/4\pi) \sim
  1.5\times 10^{52} r_{o6}^2 B_{15}^2\ergs
                                \sim 1.5\times 10^{47} r_{o6}^2
B_{12.5}^2\ergs~,
\label{eq:L}
\enq
where the suffix denotes the B-field  in gauss. The characteristic  radius
$r_o= 10^6 r_{o6}$ cm  could be the inner radius of the accretion torus
in a BH model, or, in a superpulsar model spinning with near-breakup angular
velocity, the radius of the light cylinder.
In a normal pulsar model, the field is assumed to maintain a steady value, and
the luminosity declines as the spin rate slows down. However, during the early
stages,  B might decline more rapidly than the slowing-down timescale: the power
output then declines in proportion to $B^2$. For instance, if a torus is
losing angular momentum due to magnetic torques, and is gradually draining into
the hole, the magnetic field (and therefore the associated MHD torques and
wind-driven energy losses) would gradually decline in step with the surviving
mass. Alternatively, if the central object were a rapidly-spinning neutron star,
the field may decline as uniform rotation is established and the higher moments
decay away.

    There is no reason to expect that the  field decay would  follow a power
law.  However, just as an illustration, we note, from equation (1),  that a
luminosity decay $L(t)\propto t^{-1.25}$, which would lead to $10^{47}$ erg/s
after one day, could be a consequence of a drop in B from  from $10^{15}$ G
to $3\times10^{12}$ G in a compact structure with stored energy of at least
$10^{52}$ ergs whose characteristic spin period remained constant (at a
fraction of a millisecond).

   A similar argument could be developed based on the concept of
$\alpha$-viscosity.  For a hot dense torus around the BH resulting from
collapse of the core of the progenitor star, the
$\alpha$-viscosity parameter is $\alpha\sim B^2/(4\pi\rho v_s^2) \sim  0.05
B_{15}^2 \rho_{13}^{-1}T_9^{-1}$, and the viscous accretion time for a torus of
outer radius $10^9 r_9$ cm is $t_{visc}\sim (\alpha\Omega)^{-1} \sim
% NW
10~r_9^{3/2} B_{15.5}^{-2} \rho_{13} T_9~{\rm s}$, while for a lower field of
$3\times 10^{12}$ G it is $t_{visc}\sim
1.5~r_9^{3/2} B_{12.5}^{-2} ~{\rm days}$
for $nT\sim$ constant, and the accretion of $\siml 10^{-2}\msun$ in $t\sim 1$
day is sufficient to provide a characteristic $L\sim 10^{47}\ergs$ at a day.

We envisage that the burst is triggered by the collapse of a core within a
massive stellar envelope, as in the scenario developed by Woosley and his
collaborators. A funnel along the rotation axis would have been blasted open
during the 1-100 s duration of the original burst; it would subsequently
enlarge, owing to the post-explosion expansion of the  envelope of the
progenitor star (e.g. Eichler \& Levinson 1999; Woosley 1993).
The ram pressure of the continuing MHD outflow would further enlarge the 
funnel and could, after one day, have expelled the envelope material from 
a region $10^{13} $ cm across {, even in the equatorial plane} (this 
would require velocities of no more than $10^8$ cm/s).

The magnetised wind from the compact remnant (which we assume to be relativistic)
would develop a stand-off shock before encountering the envelope material, and
shocked relativistic plasma would be deflected along the funnel walls.
% NEW
{ In the absence of magnetic fields, the contact discontinuity between the
shocked jet and  stellar gas would have a tendency to develop Kelvin-Helmholz
instabilities, which could lead to bulk heating of the stellar material.
This would extend to large Thomson depths $\tau_T \gg 1$, required to ensure
a fluid behavior leading to non-magnetic K-H instabilities. However, magnetic
fields are expected, since even if initially absent the instabilities would lead
to mixing with the highly magnetized jet material. The K-H equivalent of highly
relativistic MHD oblique shear instabilities is poorly understood, and it is
unclear how deeply , if at all, such bulk heating would penetrate the thermal 
material outside the funnel wall.  However, 
for energy deposition spread over a layer thick enough to have $\tau_T\gg 1$,
the cooling rate (due to comptonisation, bremsstrahlung and recombination) would
be high enough to reduce the temperature of the bulk-heated electrons and
protons to the equivalent black body temperature,  estimated as $T_e\sim T_p\sim
(L_m/\sigma 4\pi r_\ast^{2})^{1/4}\sim {\rm few} 10^6
L_{47}r_{13}^{-1/2}$ K.

A more efficient heating mechanism of the stellar funnel wall gas is radiative
heating, which would deposit energy within shallower layers with modest
scattering optical depth.}
% END NEW
Non-thermal electrons are expected to be accelerated behind the standoff shock 
in the jet material; the transverse magnetic field strength (which decreases as 
$1/r$ in an outflowing wind) would be of order $10^4$ G at $10^{13}$ cm -- 
strong enough to ensure that the shock-accelerated electrons cool promptly, 
yielding  a power- law continuum extending into the X-ray band. Some of these 
X-rays would escape along the funnel, but at least half (the exact proportion 
depending on the geometry and flow pattern) would irradiate the material in 
the stellar envelope.
%NEW
{ For the high radiative efficiencies expected in relativistic shocks, the
inward-directed radiative radiative heat flux would be comparable to that of a
bulk heat flux from instabilities. However since the radiative flux is deposited
in shallower layers containing less mass than the bulk heat flux, the 
radiatively heated shallow layer would be expected to be substantially hotter 
than a deeper bulk-heated region, its temperature being determined by 
photoionization equilibrium.}
% END NEW

The shocked relativistic material in the jet MHD flow exerts a pressure that is
comparable with the ram pressure of the pre-shocked  outflow.  This is
$\alpha \Lm /4\pi r^2 c$, where $r=10^{13}\r13$ is the distance of the funnel
wall, and $\alpha \siml 1$ is a geometrical factor. This must be balanced by the
thermal pressure $3 n_e kT$ in the { photoionized layers of the stellar
envelope forming} the outside the  wall of the funnel,  giving a characteristic
electron density
\beq
n_e = \alpha \Lm/6\pi r^2 c kT \sim 10^{17}\alpha L_{47} \r13^{-2}\T8^{-1}~\cmcui~
\label{eq:ne}
\enq
% NEW
{ The temperature parametrization used here is consistent with the range 
expected from photoionization equilibrium.}
% END NEW
The recombination time for hydrogenic Fe in the funnel walls photoionized by
the non-thermal continuum produced by the stand-off shock is
\beq
t_{rec}=6\times 10^{-6}\T8^{1/2}n_{17}^{-1}
    = 6 \times 10^{-6} \alpha^{-1} L_{47}^{-1}\r13^2 \T8^{3/2}~\hbox{s}~.
\label{eq:trec}
\enq
The ionization parameter is $\xi=\beta \Lm/ r^2 n_e=10^4\beta\alpha^{-1}\T8$.
In this expression $\beta < 1$ is the ratio of ionizing to MHD luminosity: it is
actually the product of two factors, namely the fraction of the MHD wind energy
that is randomised in the stand-off shocks, and the fraction of that energy
which goes into electrons that radiate in the X-ray band.  For a large fraction
of the Fe to be hydrogenic,  $\xi$ must exceed $10^{3}$, and this condition
would indeed be satisfied unless $\beta$ were very small. The effective depth
$d_i$ to which Fe and other metals can be  ionized
is given by balancing the number rate of ionizing photons per square centimetre,
$\beta\Lm/4\pi r^2 $ divided  by their mean energy (about 10 kev), against the
recombination rate $n_{Fe} \delta d_i /t_{rec}$. In this expression, $\delta
\simg 1$ takes into account the recombination of other metals besides Fe.

Standard calculations of photoionization of optically-thin slabs (e.g. Young,
1999) show that the equivalent width of the Fe K-alpha line, for solar 
abundances, is about 0.5 kev; the line is twice as strong, i.e. 1 kev equivalent 
width,  if the Fe has ten times solar abundances (and even stronger for still 
higher enrichment). These results are applicable
in the present context provided only that one further condition is satisfied:
namely, that the ionizing photons encounter a Fe
ion before being scattered by free electrons -- provided, in other words, that
$\tau_T=\sigma_T d_i n_e$ does not greatly exceed unity.  We find that
\beq
d_i \sim 3 c t_{rec} x_{Fe}^{-1} (\beta/\alpha\delta)
   \simeq 6\times 10^9 (\beta/\alpha^2\delta)
                \T8^{3/2} L_{47}^{-1}\r13^2 \zeta_{Fe\odot}^{-1}~\hbox{cm}
\label{eq:dri}
\enq
where  $x_{Fe}=4\times 10^{-5}\zeta_{Fe\odot}$ is the Fe abundance normalized to
 the
solar value.  The Thomson depth through this layer $\tau_T=\sigma_T d_i n_e
\simeq$ $4\times 10^2(\beta/\alpha\delta \zeta_{Fe\odot})\T8^{1/2} $ $\siml 3$
provided $(\beta/\alpha\delta \zeta_{Fe\odot}) \siml 10^{-2}$.   The three
parameters $\beta, \alpha$ and  $\delta$ in this expression are all somewhat
uncertain. However,  the relevant material in the funnel and cavity walls
comprises  the innermost non-collapsed layers of the precursor star so it could
well be greatly enriched in Fe (i.e. we might expect $\zeta_{Fe\odot}$ to
greatly exceed unity).

Under the foregoing conditions  the Fe K-$\alpha$ photon flux is about 0.1 of
the X-ray continuum,   and is
\beq
{\dot N}_{LFe} \sim
 10^{54} L_{47} \beta ~{\rm ph/s}
\label{eq:Lfe}
\enq

The signal observed   $t\sim 1.5$ day after the GRB 991216 burst by Piro
\etal~(2000) corresponds, for an assumed distance   $D\sim 4$ Gpc,   to
$6\times 10^{52}$ ph/s, and that observed 0.5-1 day after GRB 000214  by
Antonelli et al (2000) at $z\sim 0.47$ corresponds to $3\times 10^{51}$ ph/s.
As is clear from the above discussion, a wind
luminosity of $10^{47}$ erg/s would be sufficient to yield the observed line
signal provided that $\beta$,  the fraction of the power that goes into a
photoionizing X-ray continuum, were not below 0.06. $\beta$ is of course 
uncertain, but this number does not seem unreasonably high for the efficiency 
of particle acceleration by relativistic shocks, and suggests that our 
fiducial value of $10^{47}$ erg s$^{-1}$ for the overall luminosity need not 
be an overestimate.
The  continuum flux from GRB 991216 in the 1-10 kev band is observed to be
50-100 times stronger than the flux in the Fe line at 1.5 days.
Since this factor would only
be of order 10 for a photoionized slab with the properties envisaged here, this
suggests that most of the continuum, even in the X-ray band, could still come
from a standard afterglow model, with luminosity declining as a power-law in 
$t$, involving a decelerating blast wave. On the other hand, we cannot rule 
out the possibilility that much of the afterglow in all wavebands is due to a 
continuing power output from a compact remnant. In the latter case, the 
time-dependence could be more complicated (with possible rapid variability) and 
such effects should certainly be looked for.

\section{Discussion}

The total amount of Fe needed to explain the observed K-$\alpha$ line flux,
arising in a thin layer of the funnel walls of a collapsar model, amounts
to a very modest mass of $M_{Fe}\sim 10^{-8}\msun$, which could be Fe
synthesized in the core. The Fe-enriched core material can easily reach a
distance comparable to $r\sim 10^{13}$ cm  in 1 day for an expansion velocity
below the limit  $v\sim 10^9$ cm s$^{-1}$  inferred by Piro \etal (2000) 
(c.f. also Antonelli et al 2000) from
the line widths. Such subrelativistic velocities of the
envelope material would arise naturally from  low $\Gamma$ shocks propagating
through the star following the collapse, as well as from the burst explosion
itself at larger angles from the burst jet axis.
The natural progenitor scenario in which this can occur is a collapsar or
hypernova model (Woosley, 1993, Paczyn\'ski, 1998) with ``failed supernova"
characteristics (i.e. little or no supernova display).

The initial, energetic portion of the relativistic jet, with a typical burst
duration of $1-10$ s, will rapidly expand beyond the stellar envelope, leading
in the usual way to shocks and a decelerating blast wave. A continually
decreasing fraction of energy, such as put out by a decaying magnetar, may
continue being emitted for periods of a day or longer,  and its reprocessing
by the stellar envelope can be responsible for the observed Fe line emission
in GRB 991216. Since the energy in this tail can decay faster than $t^{-1}$,
the usual standard shock gamma-ray and afterglow scenario need not be affected,
being determined by the first 1-10 s worth of the energy input.

Finally, a few comments on how the present suggestion contrasts with other
possible interpretation of X-ray lines in the afterglow.  Because we invoke a
continuing power output after the burst (in an MHD wind) the X-ray lines that
are observed a day after the burst do not need to come from material a light day
(or more) in extent, as has been suggested earlier (Weth et al 2000, B\"ottcher,
2000).  Such a  large radius entails lower densities (and hence slower
recombination rates), and thus requires a much
larger total mass of Fe.  But the most serious problem is that it may require a
two-stage event: a supernova explosion, followed by a burst delayed by many
days, so as to allow time for a sub-relativisic Fe-rich shell to have reached
the requisite distance. Although such a model has been proposed (Stella \&
Vietri, 1998; Vietri, \etal~ 2000), detailed calculations suggest (cf B\"ottcher
\& Fryer 2000) that there are difficulties in understanding how an efficient
burst could be generated afer such a long time-delay. Such models would have
other potential observational consequences which may be used as a test, such as
a reddening and flattening of the light curve at late times as inferred in some
cases (e.g. Bloom \etal, 1999, Galama \etal, 1999).

   An alternative to the pre-ejection scenario has been proposed by B\"ottcher
\& Fryer (2000). They suggest that a very extended torus could be created during
the inward-spiralling of a compact object through the envelope of a giant or
supergiant -- a process that could be a precursor of an exotic type of
supernova. This supernova would not only generate the burst, but expel
sub-relativistic material which is shock heated when it encounters to torus. The
resultant thermal X-ray emission could display line features.  B\"ottcher \&
FryerUs suggestion is more attractive than the pre-explosion scenario, in that
(as in the model proposed here)  the X-ray lines come from a region much less
than a light day across.  However, it may not yield luminosities high enough to
explain GRB 991216. More seriously, the shocked supernova ejecta which are
postulated to emit the  X-rays have a very large optical depth, which reduces
the  equivalent width of lines and would cause the ionization edge at 9.28kev to
appear as an absorption rather than an emission feature.  The large scattering
optical depth  may even  (if it exceeds $c/v$) inhibit the escape of continuum
radiation on a dynamical timescale. However, there could be particular 
geometries in which a modified scenario along these general lines could meet 
the constraints.

     Further data on X-ray spectral features from afterglows will surely offer
important clues to the nature of the precursor star, and the compact object that
triggers the burst.

\acknowledgements{{This research has been supported by NASA NAG5-9192, the
Guggenheim Foundation, the Sackler Foundation, NSF PHY94-07194 and the Royal 
Society. We are grateful to A.C. Fabian, L. Piro and a referee for useful 
comments.}


\begin{references}

\refe{antonelli00} Antonelli, L.A., 2000, ApJL in press (astro-ph/0010221)
\refe{blo99} Bloom, J.S. \etal, 1999, Nature, 401, 453
\refe{bf00} B\"ottcher, M \& Fryer, C.L, 2000, ApJ, subm. (astro-ph/0006076)
\refe{bo99} B\"ottcher, M, 2000, ApJ in press (astro-ph/9912030)
\refe{eilev99} Eichler, D \& Levinson, A, 1999, ApJ, 521, L117
\refe{gal99} Galama, T, \etal, 2000, ApJ 536, 185
\refe{mcfad99} MacFadyen, A \& Woosley, S.E., 1999, ApJ 524, 262
\refe{mr98} \Mesz, P. \& Rees, M.J., 1998, M.N.R.A.S., 299, L10
\refe{pac98} Paczy\'nski, B., 1998, ApJ, 494, L45
\refe{piro00} Piro, L, \etal, 2000, to appear in Nov. 3 issue of Science
\refe{rm98} Rees, M.J \& \Mesz, P, 1998, ApJ, 496, L1
\refe{sm00} Sari, R. \& \Mesz, P, 2000, ApJ
\refe{tho94} Thompson, C., 1994, MNRAS, 270, 480
\refe{usov94} Usov, V.V., 1994, MNRAS, 267, 1035
\refe{vs98} Vietri, M \& Stella, L.A., 1998, ApJ 507, L45
\refe{vi00} Vietri, M, Perola, G, Piro, L \& Stella, L, 2000, MNRAS 308, L29.
\refe{whee00} Wheeler, J.C, Yi, I, Hoeflich, P \& Wang, L, 2000, ApJ in press
                                                     (astro-ph/9909293)
\refe{weth00} Weth, C, \Mesz, P, Kallman, T \& Rees, M.J, 2000, ApJ 534, 581
\refe{woo93}  Woosley, S., 1993, Ap.J., 405, 273
\refe{you99} Young, A.J., 1999, Ph.D. thesis, Cambridge University

\end{references}
\end{document}